\documentclass[12pt]{article}
\usepackage{amssymb,graphicx}
\usepackage{epstopdf}
\usepackage{amsmath,amsfonts}
\usepackage{epsfig}

\def\d{\partial}
\def\l{\left(}
\def\r{\right)}

\newcommand{\be}{\begin{equation}}
\newcommand{\ee}{\end{equation}}
\newcommand{\bea}{\begin{eqnarray}}
\newcommand{\eea}{\end{eqnarray}}
\newcommand{\bg}{\begin{gather}}
\newcommand{\eg}{\end{gather}}
\newcommand{\bseq}{\begin{subequations}}
\newcommand{\eseq}{\end{subequations}}

\begin{document}
\begin{flushright}
\end{flushright}
\vspace{10pt}

\begin{center}
  {\LARGE \bf Q-balls with scalar charge.} \\
\vspace{20pt}
A.~Levin$^{a}$, V.~Rubakov$^{b}$\\
\vspace{15pt}
  $^a$\textit{
Lomonosov Moscow State University, Physics Dept.,\\
GSP-1, Leninskie Gory, Moscow, 119991, Russian Federation
  }\\
\vspace{5pt}
$^b$\textit{
Institute for Nuclear Research of
         the Russian Academy of Sciences,\\  60th October Anniversary
  Prospect, 7a, 117312 Moscow, Russia}\\
    \end{center}
    \vspace{5pt}

\begin{abstract}
We consider Friedberg--Lee--Sirlin $Q$-balls in a 
(3+1)-dimensional model with vanishing
scalar potential of one of the fields.
The $Q$-ball is stabilized by the gradient energy of this field
and carries  scalar charge, over and beyond the global
charge. The latter property is inherent also in a model with 
the scalar potential that does not vanish in a finite field
region near the origin.
\end{abstract}

$Q$-balls of the Friedberg--Lee--Sirlin type~\cite{FRS1,FRS2,FRS3,VRbook}
exist in models with the Lagrangians of the following sort:
\be
L = {1 \over 2}(\d_\mu \phi)^2- V(\phi)
+ ( \d_\mu \chi)^{*}(\d_\mu \chi) - h \phi^2 \chi^{*} \chi \; ,
\nonumber
\ee
where $\phi$ is real scalar field whose potential $V(\phi)$ has a minimum
at $\phi=\phi_1 \neq 0$, $\chi$ is complex scalar field, and we consider
theory in 3 spatial dimensions. For large enough global charge 
$Q$ corresponding to the $U(1)$-symmetry $\chi \to \mbox{e}^{i\alpha}
\chi$, the lowest energy state is a spherical $Q$-ball with
$\phi=0$ inside and $\phi=\phi_1$ outside. Its size $R$ and energy $E$
are determined by the balance of the kinetic energy of $Q$ massless
$\chi$-quanta confined in the potential well of radius $R$ and 
the potential energy of the field $\phi$ in the interior, i.e., they are
found by
minimizing
\be
E(R) = \frac{\pi Q}{R} + \frac{4\pi}{3} R^3 V_0 \; ,
\label{1*}
\ee
where $V_0 = V(0) - V(\phi_1)$. Hence, the $Q$-ball parameters are
\be
R = \l \frac{Q}{4V_0}\r^{1/4} \; , \;\;\;\;\;
E = \frac{4\sqrt{2} \pi}{3} Q^{3/4} V_0^{1/4} \; .
\label{2*}
\ee
The $Q$-ball is stable provided its energy is smaller than the rest energy
of $Q$ massive $\chi$-quanta in the vacuum $\phi=\phi_1$,
$E(Q) < m_\chi Q$, where
\be
m_\chi = \sqrt{h} \phi_1 \; .
\label{2**}
\ee
Therefore, the estimate for the critical charge is
$Q_c \sim V_0/m_\chi^4$. At small $h$, the energy of the region where
the field $\phi$ changes from zero to $\phi_1$, which is omitted in
\eqref{1*}, is often indeed negligible even for the critical $Q$-ball.

In this note we address the question of what happens if the scalar
potential $V(\phi)$ identically vanishes,
\[
V(\phi)=0 \; ,
\]
i.e., $\phi$ is a modulus field, whose vacuum expectation value is
still non-zero,
\[
\phi_{vac} = \phi_1 > 0 \; .
\]
In similar situations in (2+1)-dimensional~\cite{Penin,Davis}
and (1+1)-dimensional~\cite{Bais} theories, the presence of a lump
gives rise to the dynamical vacuum selection: the cloud of modulus is gradually
ejected to spatial infinity, and the system relaxes to the absolute minimum
of energy (this would be the state $\phi=0$ in our case). However,
it has been pointed out~\cite{Penin,Bais} 
that the vacuum selection effect does not operate in 3 or more spatial
dimensions. 

\begin{figure}[htb!]
\begin{center}
\includegraphics[width=3.8in,height=3in]{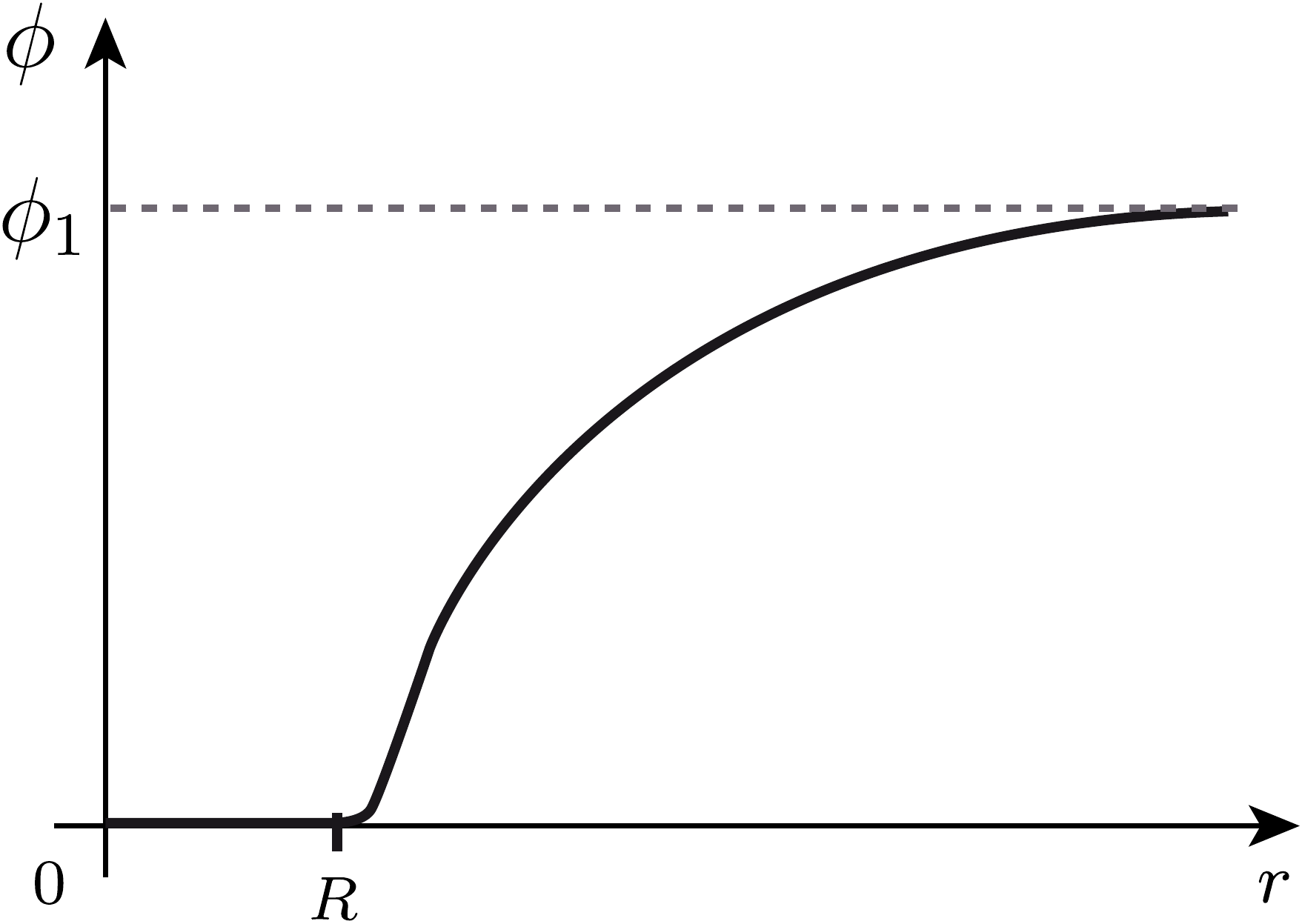}
\caption{\label{F1}
The profile of the field $\phi(r)$ in the $Q$-ball.}
\end{center}
\end{figure}

The reason for the absence of the vacuum selection is that the
$Q$-ball is stabilized by the gradient energy of
the modulus field $\phi$.
To see this, let us consider a configuraion in which $\phi(r)$ vanishes
inside a sphere of radius $R$ and gradually approaches $\phi_1$ outside this
sphere, see Fig.~\ref{F1}. We assume that the potential
well $h\phi^2(r)$ forces the wave function of $\chi$-quanta to vanish at $r>R$;
this assumption will be justified for large $Q$ in the end of the calculation. 
Then the energy of $Q$ massless $\chi$-quanta confined in the $Q$-ball
is again equal to $\pi Q/R$. The field $\phi$ is free at $r>R$, the 
minimization of its
energy gives $\Delta \phi = 0$, and the field configuration 
is\footnote{In 1 or 2 spatial dimensions, no solutions to
$\Delta \phi = 0$ would tend to the prescribed value $\phi_1$ as
$r \to \infty$. This is the basic reason for the vacuum selection
in these dimensions.} 
\be
\phi (r) = - {C \over r} + \phi_1 \; , \;\;\;\;\;\; r>R \; ,
\label{3}
\ee
where 
\be
C= R \phi_1 \; .
\label{4X}
\ee
The parameter $C$ is natuarally 
interpreted as the scalar charge of the $Q$-ball.
Hence, the $Q$-balls we discuss experience long-ranged attraction
mediated by the modulus field $\phi$. 

The gradient energy of the $Q$-ball hair \eqref{3} is
\be
E_{\phi} 
= \int_{R}^{\infty} 
\frac{1}{2} \left( {C \over r^2} \right)^2  4\pi r^2 dr = 2\pi {\phi_1}^2 R \; .
\label{add1}
\ee
Hence, the total energy of the system is given by
\be
E (R) =\pi {Q \over R} + 2\pi {\phi_1}^2 R .
\ee
By minimizing this expression with respect to $R$, we obtain the size,
energy and scalar charge of the $Q$-ball,
\bea
R&=& \frac{\sqrt Q} {\sqrt{2} \phi_1}
\label{4+}
\\
E &=& 2 \sqrt{2} \pi \phi_1 \sqrt Q
\label{11}
\\
C &=& \phi_1 R = \sqrt {\frac{Q}{2}}
\label{12}
\eea
Note that the dependence of $R$ and $E$ on $Q$ is entirely different
from \eqref{2*}. Parameters of the critical $Q$-ball are obtained
by equating the energy \eqref{11} to the rest energy of 
$Q$ quanta of the field $\chi$ in the vacuum
$\phi=\phi_1$, i.e., $E(Q_c) \sim m_\chi Q_c$,  where $m_\chi$ is still
given by \eqref{2**}. We obtain
\be
Q_c \sim \frac{8\pi^2}{h} \; , \;\;\;\;\;\;
R_c \sim \frac{2\pi}{m_\chi} \; , \;\;\;\;\;\;
E_c \sim \frac{8 \pi^2}{h} m_\chi \; .
\label{5*}
\ee
We emphasize that these expressions are estimates only, since the critical size
$R_c$ is of the order of the $\chi$-boson mass in vacuum $\phi=\phi_1$,
so our approximation of vanishing $\chi$-boson wave function
at $r>R$ is not valid for the critical $Q$-ball.

This approximation {\it is} valid for $Q\gg Q_c$, so the expressions
\eqref{4+}, \eqref{11}, \eqref{12} are exact in the large-$Q$ limit.
To see this, let us estimate the actual spatial extent of  the
$\chi$-boson wave function in the region $r>R$. There, the wave function
obeys
\[
\frac{1}{r^2} \frac{d}{d r} \l r^2 \frac{d \chi}{dr} \r
+ \omega^2 \chi - m_\chi^2 \l 1 - \frac{R}{r} \r^2 \chi = 0 \; ,
\]
where $\omega = \pi/R$ is the $\chi$-quantum energy. It is legitimate to
neglect the
second term, so the WKB solution is
\[
\chi \propto \mbox{exp} \left[ -\int_R^r ~dr~
m_\chi \l 1 - \frac{R}{r} \r \right] \; .
\]
For $r-R \ll R$ this gives
\[
\chi \propto \mbox{exp} \left[ - m_\chi \frac{(r-R)^2}{2R} \right] \; .
\]
Hence, the spatial extent is of order $\Delta r \sim \sqrt{R/m_\chi}$.
For $Q\gg Q_c$ we have $R \gg m_\chi^{-1}$ 
and hence $\Delta r \ll R$, as promised.

\begin{figure}[htb!]
\begin{center}
\includegraphics[width=3.8in,height=3in]{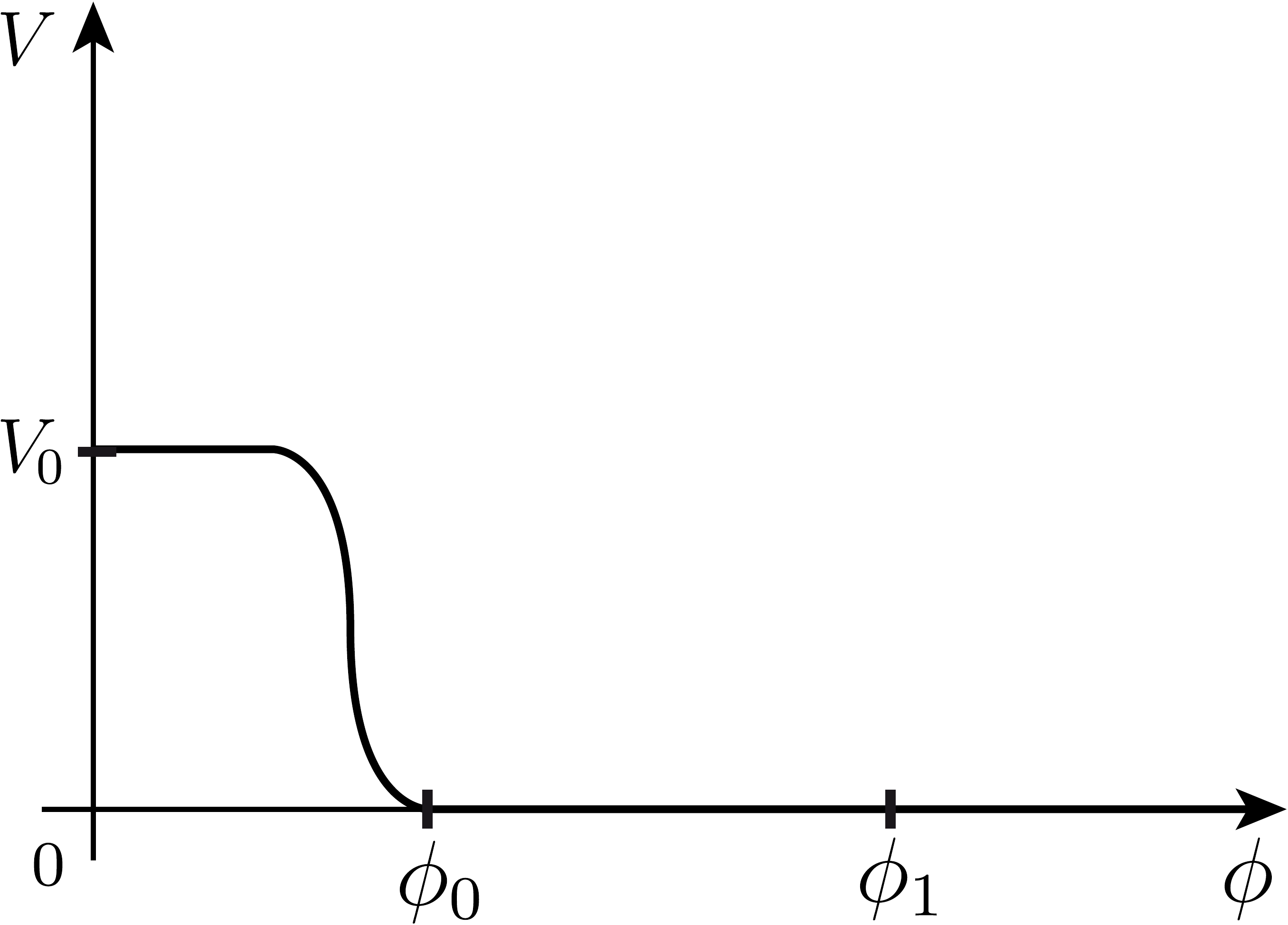}
\caption{\label{F2}
Scalar potential with hump near the origin.
}
\end{center}
\end{figure}

For completeness, let us consider a model in which the potential
$V(\phi)$ vanishes at $\phi>\phi_0$  but is non-zero
at $\phi<\phi_0$, as shown in Fig.~\ref{F2}.
An easily tractable case is the vacuum $\phi_1 \gg \phi_0$.
In that case, there may exist a range of values of the global charge
$Q$ in which the $Q$-ball properties are still governed by
the gradient energy of the field $\phi (r)$ rather than its potential
energy. This occurs when $V_0 R^3 \ll \phi_1^2 R$, see \eqref{add1}.
We recall the result \eqref{4+} and find that the gradient energy
dominates over the potential energy for $Q\ll \phi_1^4/V_0$.
The range of global charges in question
is not empty if $\phi_1^4/V_0 \gg Q_c$,
i.e., $V_0 \gg m_\chi^2 \phi_1^2$. If so, then
in the intermediate range of global charges,
\[
\frac{8\pi^2}{h} \ll Q \ll \frac{\phi_1^4}{V_0} \; ,
\]
$Q$-balls have the properties \eqref{4+}, \eqref{11}, while for
$Q\gg \phi_1^4/V_0$ we are back to \eqref{2*}. 
For $Q \sim \phi^4 / V_0$ the potential and gradient energies are of
the same order, and again by minimizing the energy with respect
to $R$ we obtain
\be
R^2= \frac{1}{4 V_0} \l \sqrt {4 Q V_0 + \phi_1^4} - \phi_1^2\r
\; .
\label{aug19-1}
\ee
It is worth noting that in either
case the field profile at $r > R$ is given by
\eqref{3}, i.e., the $Q$-ball carries non-vanishing scalar charge
$C=\phi_1 R$, where $R$ is given by either  \eqref{4+} or \eqref{2*}
or \eqref{aug19-1}.

To summarize, $Q$-balls in models with $V(\phi)$ vanishing away from
the origin are rather different from the usual Friedberg--Lee--Sirlin
$Q$-balls, since they are stabilized by gradient, rather than potential
energy. Also, they carry Coulomb-like scalar hair. In this respect they
are similar to the BPS monopoles. Unlike the BPS monopoles,
however, the $Q$-balls experience long-range interactions between
themselves: there is no other force to counterbalance the
attraction due to the massless scalar field $\phi$.

\end{document}